\documentclass[sigconf]{acmart}
\usepackage{siunitx}

\title{Toxic Bias: Perspective API Misreads German as More Toxic}

 \author{Gianluca Nogara}
 \authornote{These authors contributed equally to this research.}
 \email{gianluca.nogara@supsi.ch}
 \affiliation{
 \institution{University of Applied Sciences and Arts of Southern Switzerland}
 \country{Switzerland}}
 
 \author{Francesco Pierri}
 \authornotemark[1]
 \affiliation{
 \institution{Politecnico di Milano}
 \country{Italy}}
 
 \author[ ]{Stefano Cresci}
 \affiliation{
 \institution{IIT-CNR}
 \country{Italy}}
 
  \author[ ]{Luca Luceri}
 \affiliation{
 \institution{USC Information Sciences Institute}
 \country{USA}}

\author{Petter Törnberg}
 \affiliation{
 \institution{University of Amsterdam}
 \country{Netherlands}}
 
 \author[ ]{Silvia Giordano}
 \affiliation{
 \institution{University of Applied Sciences and Arts of Southern Switzerland}
 \country{Switzerland}}
 
\renewcommand{\shortauthors}{Nogara, et al.}

\begin{document}

\begin{abstract}
{\color{red}\textbf{Please check and cite the published version of this paper in the Proceedings of the 19th AAAI International Conference on Web and Social Media (ICWSM'25).}}\\
Proprietary public APIs play a crucial and growing role as research tools among social scientists. Among such APIs, Google's machine learning-based Perspective API is extensively utilized for assessing the toxicity of social media messages, providing both an important resource for researchers and automatic content moderation. However, this paper exposes an important bias in Perspective API concerning German language text. Through an in-depth examination of several datasets, we uncover intrinsic language biases within the multilingual model of Perspective API. We find that the toxicity assessment of German content produces significantly higher toxicity levels than other languages. This finding is robust across various translations, topics, and data sources, and has significant consequences for both research and moderation strategies that rely on Perspective API. For instance, we show that, on average, four times more tweets and users would be moderated when using the German language compared to their English translation. Our findings point to broader risks associated with the widespread use of proprietary APIs within the computational social sciences.

\end{abstract}

\maketitle


\section{Introduction}
In recent years, abusive interactions have become a fixture of the online experience, with approximately 40\% of U.S. adults self-reporting as victims of online abuse \cite{vogels2021state, pierri2020diffusion}. The prevalence of such language on social media has become a major concern for both researchers and the public, often under the label of online ``toxicity". Currently, the leading approach to detecting toxicity is the Perspective API\footnote{\url{https://www.perspectiveapi.com/}}, offered by the Jigsaw unit of Google \cite{lees2022new} and designed to ``help mitigate toxicity and ensure healthy dialogue online". The API is widely used both within academic research efforts and for content moderation on news and social media platforms. As of January 2024, there are over 1,400 documents on Google Scholar mentioning the term ``Perspective API", and the API is used by major platforms such as Reddit, The New York Times, El Pais, and Faceit.

Perspective API's machine learning model is developed through supervised learning, utilizing a vast dataset of millions of comments gathered from diverse online platforms, including forums like Wikipedia and The New York Times, and encompassing more than 20 languages. Perspective API defines \textit{toxic} messages as texts that use ``rude, disrespectful, or unreasonable language that is likely to make someone leave a discussion" \cite{Gehman2020RealToxicityPromptsEN,Welbl2021ChallengesID}. The API provides several different toxicity scores, namely \textit{Severe toxicity, Insult, Profanity, Identity attack, Threat}, and \textit{Sexually explicit}. In a 2019 SemEval Task, Perspective API has been shown to perform better than other transformer-based models on canonical datasets for hate speech detection \cite{pavlopoulos-etal-2019-convai}. More recent work found that its scores differ from human labels \cite{Welbl2021ChallengesID}. The toxicity score, which ranges between 0 and 1, does not have a meaning in absolute terms, and the traditional approach to identify toxic or hateful content in online social platforms involves defining a threshold on the toxicity score, which is generally in the region of 0.5-0.7 \cite{hua2020towards,hua2020characterizing}, above which content is considered to be toxic.

While technologies such as Perspective API promise a safer, more respectful digital environment, their effectiveness is contingent on their accuracy and impartiality across diverse languages and cultural contexts. This paper aims to address the following research question:
\begin{itemize}
\item \textbf{RQ: Is Perspective API biased towards the German language?}
\end{itemize}
By leveraging two large-scale datasets of multilingual Twitter conversations and random Wikipedia summaries, we identify and report on a systemic bias within Perspective API when dealing with German texts, showing a tendency of the tool to assign higher toxicity scores to German content compared to other languages. We show that toxicity scores for texts in German are consistently significantly higher than those for any other language, \textit{regardless of the content}. Our findings suggest a strong negative bias against the German language, potentially due to artifacts or biases in the training dataset or the model itself, which is challenging to investigate further given the black-box nature of the tool. Our findings not only question the reliability of a tool employed by several large online platforms -- potentially leading to unjust censorship or misrepresentation in online spaces -- but also aggravate the growing reliance on proprietary, black-boxed APIs like Perspective within the academic realm. These systems present a fundamental problem as researchers and users cannot scrutinize or understand the inner workings of the algorithms, leading to a substantial risk of systemic biases being embedded and perpetuated through their employment across news outlets, social media, and various online platforms.

The outline of this paper is the following: we first examine recent work employing Perspective API for toxicity research; next, we describe the datasets employed in our analysis; then, we show results for a number of different analyses. Finally, we outline our findings and discuss implications, limitations and future directions.

\section{Perspective API and toxicity research}
Perspective API has become a foundation for academic research on online abuse and incivility~\cite{rieder2021fabrics}, as well as a relevant case study for legal and social scholars interested in the regulatory nuances of socially relevant AI systems~\cite{friedl2023dis}. A diverse and rich research field relies on the capacity of the API to reliably assess these aspects of online text. Rather than reviewing the diverse contributions that employ Perspective API to study online toxicity, we will here examine some examples of research that draws on the API concerning the analysis of German language messages, to assess the potential implications of the findings of this paper.

There are many examples of research using Perspective API for cross-national comparison of toxic communication on social media. Rye et al. \cite{rye2020reading}, for instance, analyze 14 months of data activity on the far-right platform Dissenter, using Perspective API to measure the toxicity of messages and derive the propensity of specific content to elicit toxic comments. They find that ``deutschland.de" is the second most toxic domain, with most angry comments being related to the Muslim diaspora in Europe. Similarly, Hoseini et al. \cite{hoseini2023globalization}, analyze Telegram messages related to the QAnon conspiracy shared between September 2019 and March 2021, finding that German and Portuguese messages are more toxic than those in English.

Even when studies are not explicitly comparative, they often include cross-language analysis of toxicity, which may potentially be vulnerable to the type of model biases discussed in this paper. Smirnov et al. \cite{smirnov2023toxic}, for instance, study the effect of toxic comments directed to Wikipedia's volunteer editors. Analyzing over 57 million comments on the six most active language editions of Wikipedia (English, German, French, Spanish, Italian, and Russian), they find that toxic speech could lead to a substantial loss of productivity among editors. Differently from the majority of other contributions, Ghosh Roy et al.~\cite{ghosh2020leveraging} do not directly use Perspective API to identify toxic tweets, but rather use toxicity scores from the API as features in a transformers-based hate speech classifier for English, German, and Hindi. This usage, however, could leak Perspective API's biases into the trained classifier.

Many studies focus specifically on German-language debates, seeking to draw out the dynamics of toxicity within these social spaces. Lasser et al. \cite{lasser2023collective} study more than 1M tweets posted on Twitter from 2015 to 2018 in German, a period characterized by a ``large influx of refugees into Germany which sparked heated discussions and political polarization". They measure the interaction between incivility (hatefulness and toxicity) and extremity (of speech and speakers) with other dimensions of discourse. For this purpose, they use Perspective API to measure the overall toxicity score as the average of five attributes (toxicity, severe toxicity, profanity, insult, and identity attack). They report very high values of average toxicity and highlight which behaviors can reduce hate speech. Another example is Czymara et al.'s analysis of the eﬀect of Islamist terror attacks on ethnic insulting comments on YouTube \cite{czymara2023catalyst}, showing how terror attacks boost interest in immigration-related topics in general, and lead to a disproportional increase in hate speech in particular. Focusing on the ``Identity attack" attribute of Perspective API, they classify insulting comments as those with a score surpassing 0.9. The study reveals varying proportions of ethnically insulting comments, ranging from 2.7\% in the French data to 5.4\% in the German data and 6.3\% in the UK data.

While Perspective API has seen widespread use, few studies have systematically evaluated the accuracy and reliability of the model either across languages or specifically in German. Among them,~\cite{gargee2022analyzing} highlights a discrepancy in Perspective API's toxicity assessments of sentences that are semantically equivalent, but with different arrangements of words. In detail, they show that switching from active to passive voice, as well as reordering words in a sentence, could lead to significantly different toxicity scores. Then, Wich et al. study toxic German-language speech on Telegram~\cite{wich2022introducing}. They develop a custom classifier and compare its performance with Perspective API, finding that their model exhibits a higher F1 score ($54.75\%$ vs. $53.50\%$) and macro F1 score ($73.20\%$ vs. $70.51\%$). Mihaljevic et al. \cite{mihaljevic2023toxic} use antisemitism in the German language on Telegram to examine how Perspective API scores correlate with different sub-forms of antisemitism. They also demonstrate that manipulating messages with widely known antisemitic codes can substantially reduce toxicity scores, making it easy to bypass content moderation based on Perspective API. While such studies suggest the need for more research on potential biases of the model, the black-box nature of the model limits the capacity of researchers to understand the limitations and dynamics of the classification. Research therefore has to draw on observational data to seek to identify biases in model classification.

\section{Data}
\label{sectionData}
Our analyses are based on a combination of real-world data from Twitter and Wikipedia, as described in the following subsections.

\subsection{Dataset 1: Representative sample of international tweets across languages}

\begin{table}[!]
\centering
\begin{tabular}{lll}
Austria & Belgium & France \\
 \hline
Germany & Ireland & Italy \\
 \hline
Netherlands & Poland & Spain \\
 \hline
Sweden & Switzerland \\ \hline
\end{tabular}
\caption{Countries included in the Dataset 1 (representative sample of international tweets across languages).}
\label{tab:countries-dataset-one}
\end{table}

Our first Twitter dataset contains a random sample of tweets shared between 2018 and 2021 in 14 countries (see Table\ref{tab:countries-dataset-one}), collected using the Twitter Academic API on Sept. 12, 2022. The collection was done by repeatedly selecting a random time point between 1/1/2018 and 1/1/2022, and fetching the 500 most recent messages sent in a particular country with a particular language, until approximately \num{250000} messages were collected for each region. The country and language were filtered using the Twitter API \textit{place\_country} parameter, combined with the \textit{lang} parameter to specify the language. The resulting dataset represents a roughly representative sample of Twitter messages uniformly distributed over time. The procedure thus produces a dataset with a pseudo-random sample of Twitter messages from 11 countries over the specified time frame. The total number of tweets equals \num{2666696}, while those shared in German-speaking countries, such as Austria (AT), Germany (DE), and Switzerland (CH), are \num{733451} (27.5\%). Having multiple countries and languages enables showing the unique biases of German classification, and including several German-speaking countries allows verifying that the results do not follow from the dynamics of online conversation in a specific country.

\subsection{Dataset 2: COVID-19 tweets in Italian and German}
\label{subsecCovid-datasets}
The second Twitter dataset contains tweets related to COVID-19 vaccines in the Italian and German languages, shared between November 1, 2020, and June 30, 2021 -- a time when vaccination campaigns kicked off in many European countries. The resulting number of tweets is \num{1453450} for Italian and \num{1219445} for German. The data were collected through Twitter's Historical Search API in July 2021 and are described in detail in \cite{giovanni2022vaccineu}. These larger-scale data enable a deeper comparison of Perspective API's labeling dynamics between German and another language.

\begin{figure}[t]
    \centering
    \includegraphics[width=\linewidth]{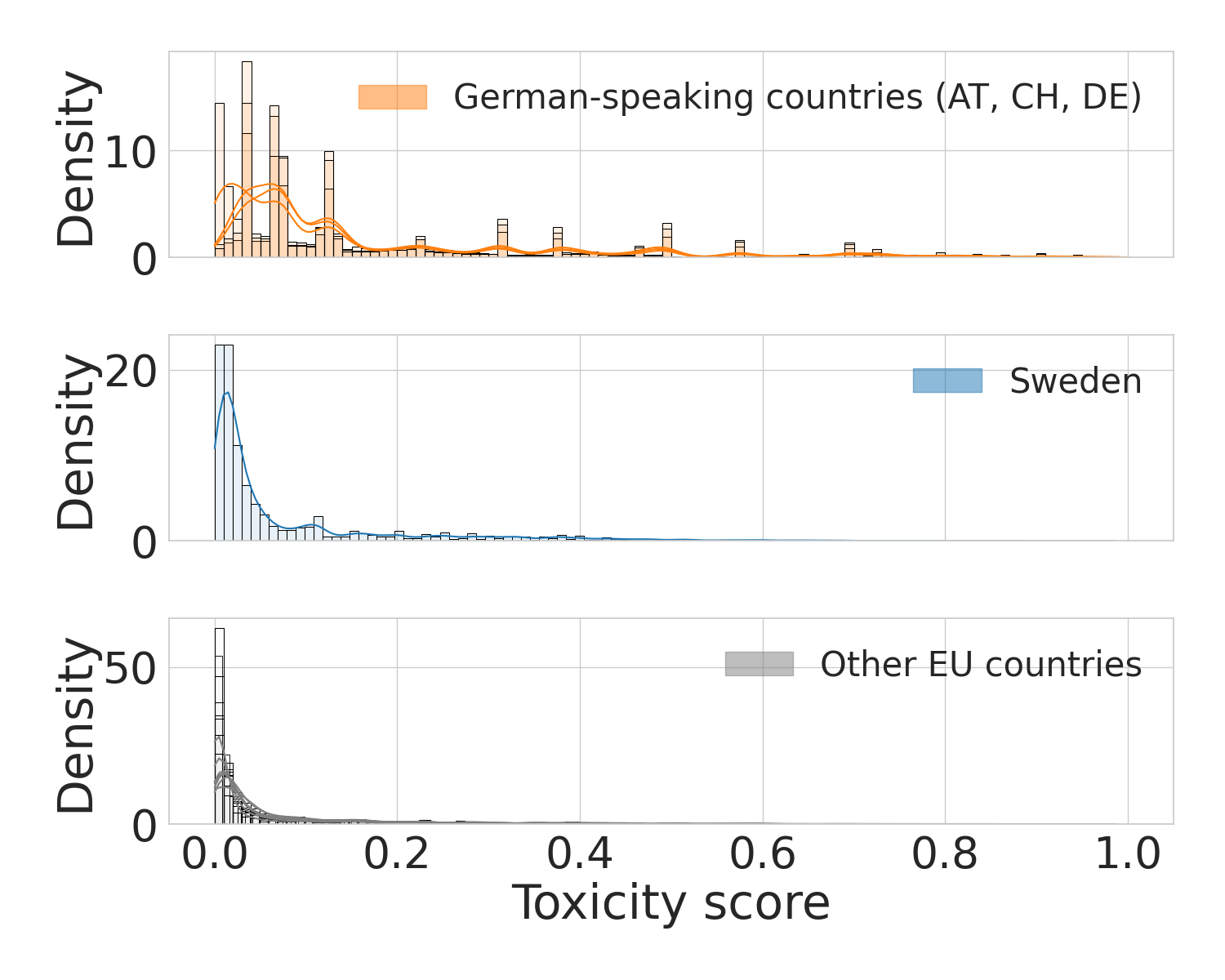}
    \caption{Distribution of toxicity scores for tweets shared in German-speaking countries (Austria, Switzerland, and Germany) versus those in other EU countries, from Dataset 1. Histograms are built with 100 equal-width bins. Distributions are statistically different at $\alpha=0.05$ according to a Kruskal-Wallis test. The median toxicity of tweets shared in German-speaking countries is 0.075, and $\sim0.023$ when considering the overall distribution of tweets in other EU countries.}\label{fig:de_eu_bias}
\end{figure}

\subsection{Dataset 3: Random sample of Wikipedia texts}
The third dataset consists of a random sample of summaries from Wikipedia pages in English and German language. Specifically, we collected \num{12279} summaries of a random sample of pages available in both languages leveraging Wikipedia Python library\footnote{\url{https://pypi.org/project/wikipedia/}}. We further collected a similar amount of random texts in Arabic, Chinese and Japanese\footnote{They do not correspond to the same texts across the five languages, as we were unable to retrieve summaries that satisfied this condition with the Wikipedia API.} to test whether languages with non-ASCII characters might exhibit biases in toxicity scores. Including random texts allows verifying that the biases identified do not apply only to Twitter messages but to any text.

\begin{figure}[!t]
    \centering
    \includegraphics[width=\linewidth]{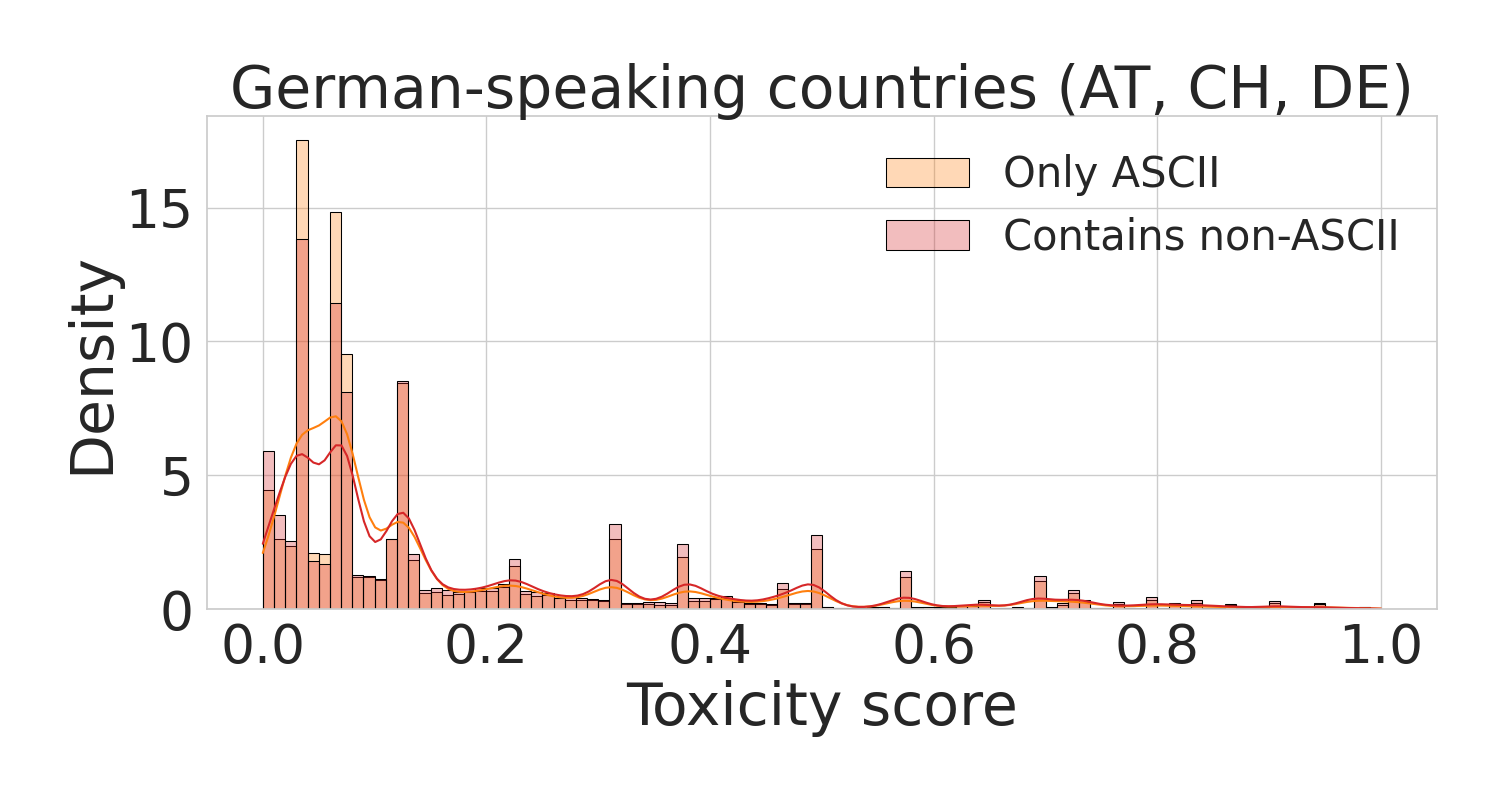}
    \caption{Distribution of toxicity scores for tweets shared in German-speaking countries (Austria, Switzerland, and Germany) from Dataset 1, separating texts that contain at least one non-ASCII character from those that only contain ASCII characters. Histograms are built with 100 equal-width bins. The median toxicity of the two distributions are respectively 0.08 and 0.07.}\label{fig:de_ascii}
\end{figure}

\begin{figure}[!t]
    \centering
    \includegraphics[width=\linewidth]{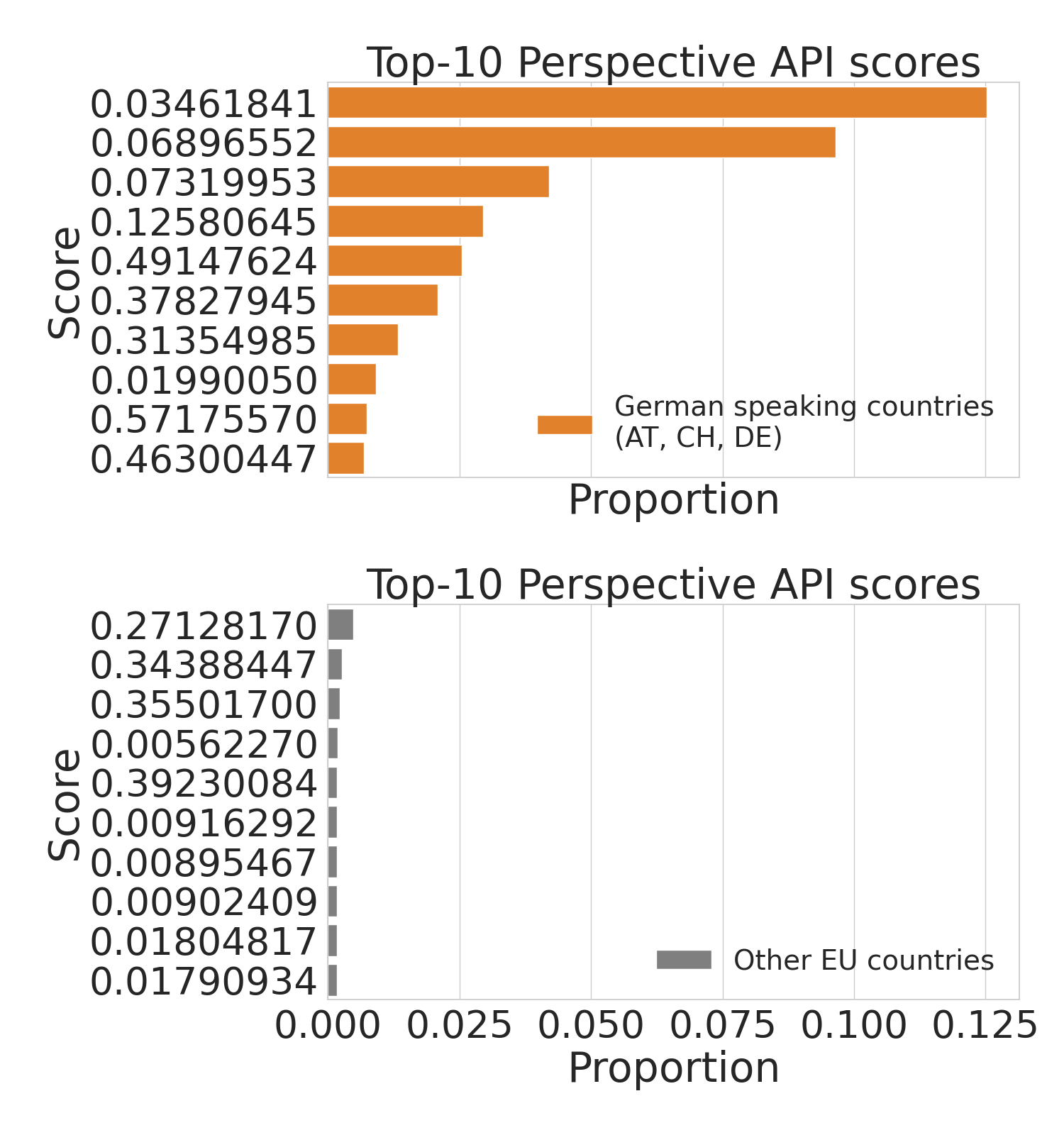}
    \caption{Top 10 most frequent Perspective API scores rounded at 8 decimal digits for tweets shared in German-speaking countries (Austria, Switzerland and Germany) and those in other EU countries, from Dataset 1. }\label{fig:de_eu_bias_digits}
\end{figure}

\section{Analyses and Results}
This section presents the results of the analyses conducted on the three aforementioned datasets. In each case, we ran Perspective API on all available texts focusing on the attribute ``Toxicity" -- the estimated probability for a comment to be perceived as toxic -- as it is the most general and widely employed in the literature. Specifically, we employed Perspective API in September 2022 to obtain toxicity scores for Dataset 1, whereas we ran it between September and December 2023 for labeling Datasets 2 and 3.

Our analyses first investigate the distribution of toxicity scores in a random sample of multilingual tweets (Dataset 1). Next, we perform a similar analysis for tweets in Italian and German language related to COVID-19 vaccines (Dataset 2). For this dataset, we also assess the potential impact of Perspective API's language bias when it is used for automatically flagging toxic messages for removal -- a common content moderation strategy \cite{rieder2021fabrics}. Lastly, we analyze biases in Perspective API scores when considering a random set of Wikipedia summaries in their corresponding English and German language versions (Dataset 3).

\begin{figure}[!t]
    \centering
\includegraphics[width=\linewidth]{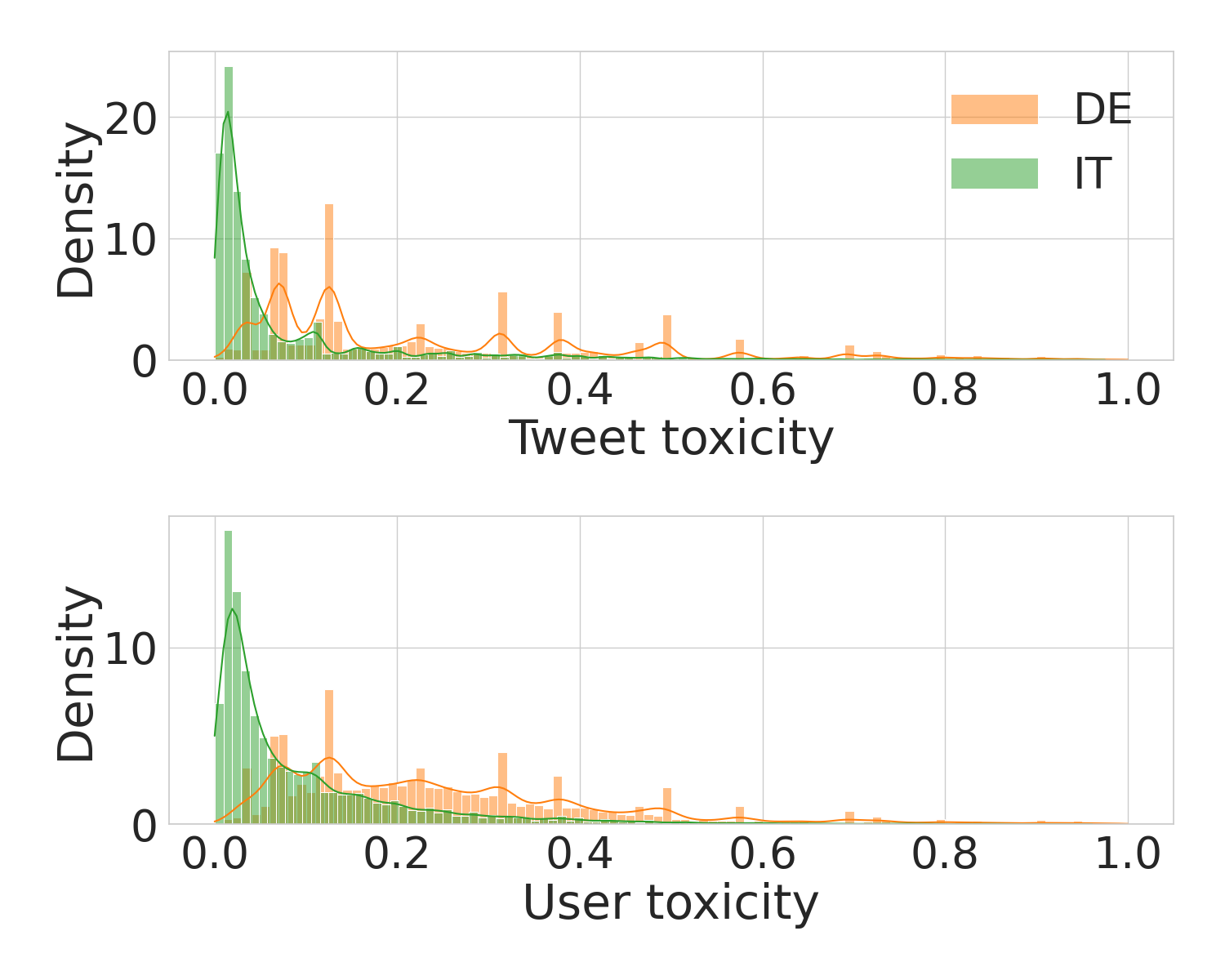}
\caption{Distribution of toxicity scores for COVID-19 vaccine-related tweets (top) and users (bottom) for German and Italian language, from Dataset 2. Histograms are built with 100 equal-width bins. Median values for tweet toxicity are DE = 0.132, IT = 0.026; for user toxicity DE = 0.212, IT = 0.048.}\label{fig:de_it_tweets_user}
\end{figure}

\subsection{German tweets are labeled as more toxic than those from other languages}
We first study the distribution of toxicity scores for tweets shared in German-speaking countries -- namely Austria, Germany, and Switzerland -- compared to other countries. As shown in Figure \ref{fig:de_eu_bias}, tweets from German-speaking countries (which are mostly in the German language) are significantly more toxic (Kruskal-Wallis $P < .001$) than those from other countries, with a median toxicity of 0.075 versus 0.023 when considering tweets from the two groups of countries altogether. The latter distributions closely resemble smooth exponential distributions, which is typical of a range of phenomena on social media \cite{adamic2001search,kwak2010twitter}, whereas the former are spiky -- suggesting the presence of some type of error or classification artifact.

\subsubsection{Non-ASCII characters are not related to higher toxicity of tweets}
To investigate whether the presence of non-Latin characters (such as ä, ö, ü) might cause this abnormal behavior, we also plot in Figure \ref{fig:de_eu_bias} the distribution of tweets shared in Sweden separately from other countries (see middle panel), finding no evidence of spikes. We further compare the toxicity of texts in German that contain at least one non-Latin character (i.e., with an ASCII code greater than 127) from those that only contain Latin characters, finding very similar spiky distributions in the two classes (see Figure \ref{fig:de_ascii}). We observe similar patterns also when separating tweets on the number of non-ASCII characters they contain in the range $[1, 12]$ (i.e., up to the 99th percentile of the number of non-ASCII characters in tweets from German-speaking countries). We provide the figure in the Appendix (Figure \ref{fig:de_ascii_appendix}).

\subsubsection{Abnormal peaks of toxicity score in German tweets}
To explore abnormal peaks of toxicity, we look into the top 10 most frequent toxicity scores returned by the API after rounding them at 8 decimal digits. As shown in Figure \ref{fig:de_eu_bias_digits}, tweets in German exhibit an unusual pattern, as some very specific scores are extremely common. While a smooth distribution would imply that each score should account for a negligible proportion of all scores, as in the case of the distribution for all other languages (in grey in Figure \ref{fig:de_eu_bias_digits}), a few scores are disproportionately frequent for German texts -- in line with abnormal spikes highlighted in Figures \ref{fig:de_eu_bias} and \ref{fig:de_ascii}, accounting for more than 10\% of the total toxicity scores returned by the Perspective API. This offers further evidence of the presence of some internal issues in the toxicity model leading to erroneous labeling of German language text.

\subsubsection{No evidence of specific words associated with abnormal toxicity scores}
Notwithstanding the fact that Perspective API is a black-box machine learning model, we perform an exploratory analysis to understand whether specific words might cause the artifacts in the toxicity score. Specifically, we first assign tweets to two classes: \textit{Abnormal} tweets if they are assigned one of the scores described in the previous section, \textit{Normal} otherwise. Then, we compute TF-IDF embeddings of such tweets (excluding words that appear less than 1\% of the time), and we train and test two off-the-shelf machine learning classifiers (Random Forest and Gaussian Naive Bayes) using the \textit{scikit-learn} \cite{pedregosa2011scikit} library in the binary classification task of predicting whether a tweet belong to one of the two classes. We observe that they perform no better than a random baseline across several performance metrics, as shown in Figure \ref{fig:classification} of the Appendix. We also leverage the \textit{Fightin' Words} approach first described in \cite{monroe2008fightin} and recently employed by \cite{gligoric2024nlp} to measure statistically significant differences in words between the two classes, which might be causing the spikes in the distribution. By looking at the top-10 differentiating words for the Abnormal class, as shown in Figure \ref{fig:fighting_words} of the Appendix, we observe that they are associated with a slightly larger toxicity score and correspond to very common terms (e.g., \textit{nicht}, \textit{die} and \textit{sie}). However, tweets that do not contain such words still exhibit the abnormal peaks.

\subsection{German vaccine tweets are more prone to removal}
\label{subsecCovid-results}
To examine the implications of the identified biases in toxicity scores for realistic real-world scenarios of content moderation, we leverage the dataset on COVID-19 vaccine-related tweets. COVID-19 information has frequently been raised as an example in which content moderation is particularly crucial \cite{ferrara2020misinformation,krishnan2021research}.

\begin{figure}[t]
    \centering
\includegraphics[width=\linewidth]{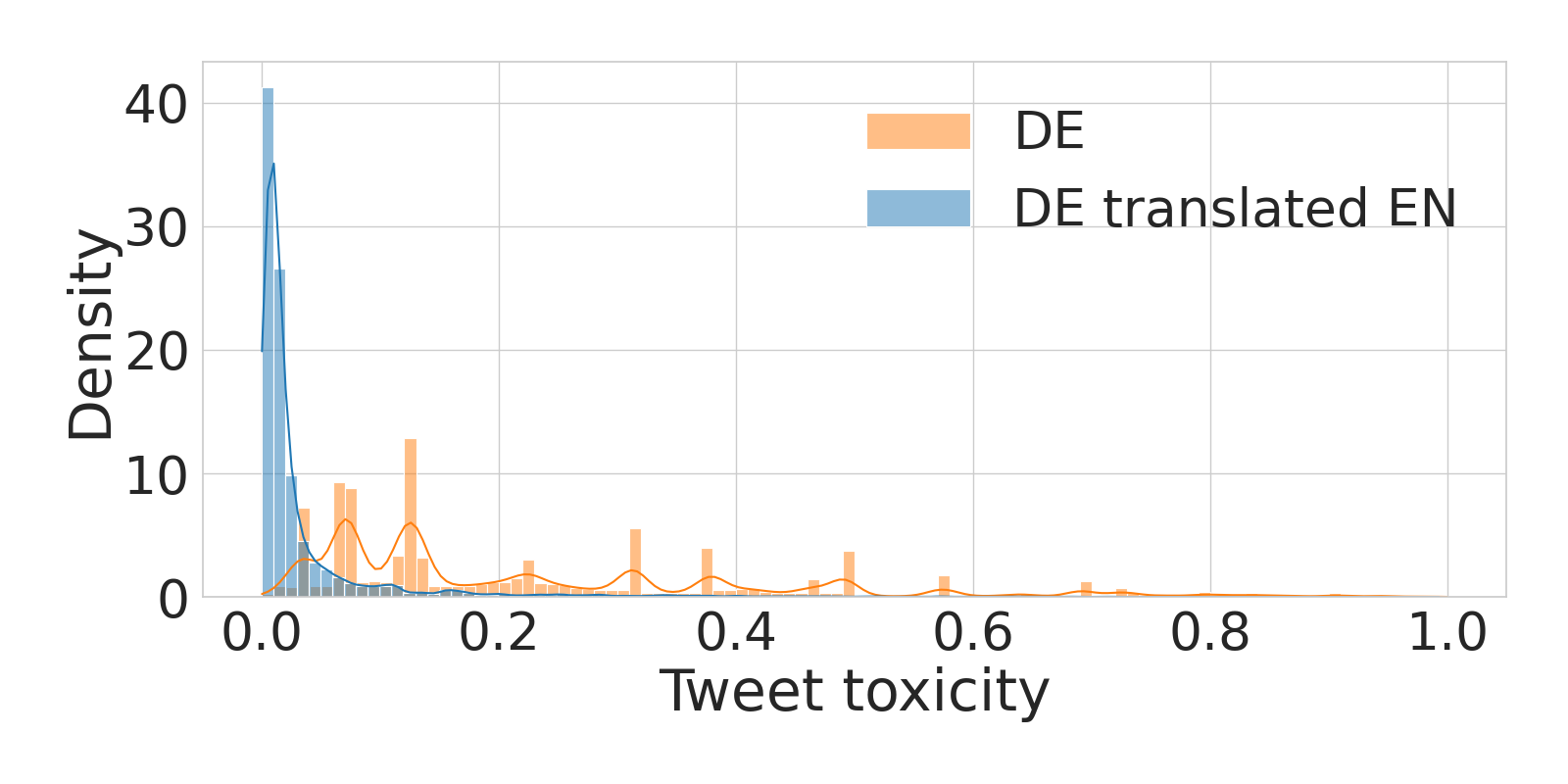}
\caption{Distribution of toxicity scores for German COVID-19 vaccine-related tweets and for their English translation, from Dataset 2. Histograms are built with 100 equal-width bins. Median values are: DE = 0.132, EN = 0.012.}\label{fig:de_translated_tweets}
\end{figure}

\subsubsection{Toxicity of tweets and users}
We first analyze the distribution of toxicity scores for tweets shared in German and Italian. As shown in the top panel of Figure \ref{fig:de_it_tweets_user}, German tweets are significantly more toxic than Italian tweets. The bottom panel of Figure \ref{fig:de_it_tweets_user} highlights that the same result also holds when considering user-level toxicity, which we computed as the average of all tweets shared by each user. The Italian and German distributions are statistically different in both cases according to a Kruskal-Wallis test ($P < .001$).

\subsubsection{Toxicity of tweets translated to English}
To rule out the possibility that the higher toxicity scores assigned to German texts are due to more toxic content, we translate both German and Italian tweets into English and we subsequently use Perspective API to compute toxicity scores for the English texts. We leverage Argos Translate\footnote{\url{http://www.argosopentech.com/}} to obtain English translations of our tweets.
Figure \ref{fig:de_translated_tweets} shows the distribution of toxicity scores for all German tweets (solid line) and their English translation equivalents (dashed line). We observe that the tweets translated in English are much less toxic, with their distribution being similar to that of generic tweets in non-German languages, presented in Figure \ref{fig:de_eu_bias}.

\begin{figure}[t]
    \centering
\includegraphics[width=\linewidth]{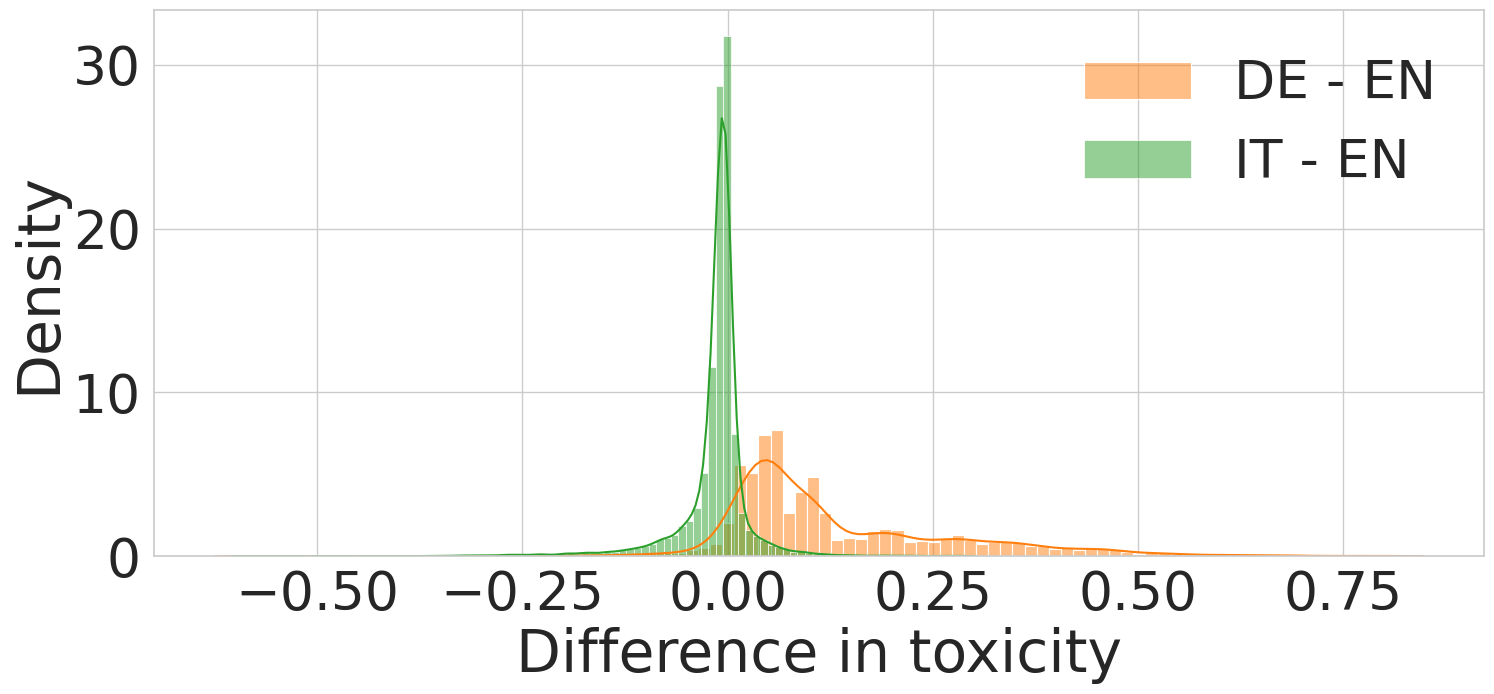}
\caption{Distribution of the differences in toxicity scores for German and Italian tweets compared to their English translation, using a random sample of \num{10000} tweets from Dataset 2. Histograms are built with 100 equal-width bins. Median values are DE = +0.085, IT = -0.006.}\label{fig:de_it_translated_difference}
\end{figure}

However, it may be that the machine-translation will tend to reduce the toxicity of messages, for instance by avoiding to translate certain swearwords or slang. To examine whether this is the case, we also compute the difference in toxicity between a random sample of \num{10000} tweets in German compared to their respective English and Italian translation. Figure \ref{fig:de_it_translated_difference} shows the distribution of these differences. We observe that, on average, German messages are 8.5\% more toxic (i.e., median difference = +0.085) than the corresponding English messages, as also shown by the skewed orange-colored distribution. Contrarily, the differences between Italian and English-translated messages appear to follow a normal distribution centered around 0 and, in fact, Italian messages only show negligible differences (median difference = 0.01) when compared to their English translations. These results unequivocally demonstrate that the larger toxicity scores for German texts do not depend on the content of such texts, but are rather the consequence of an inherent language bias in Perspective API. To rule out potential biases in the translation that might modify the toxicity, we perform the same experiments by starting from a random sample of \num{10000} Italian tweets and translating them to German and English to compute differences. Also in this case, we obtained similar results, but omit the figure for the sake of brevity.

\subsection{Correlation between toxicity and difference with English translation}
\begin{figure}[t!]
    \centering
\includegraphics[width=\linewidth]{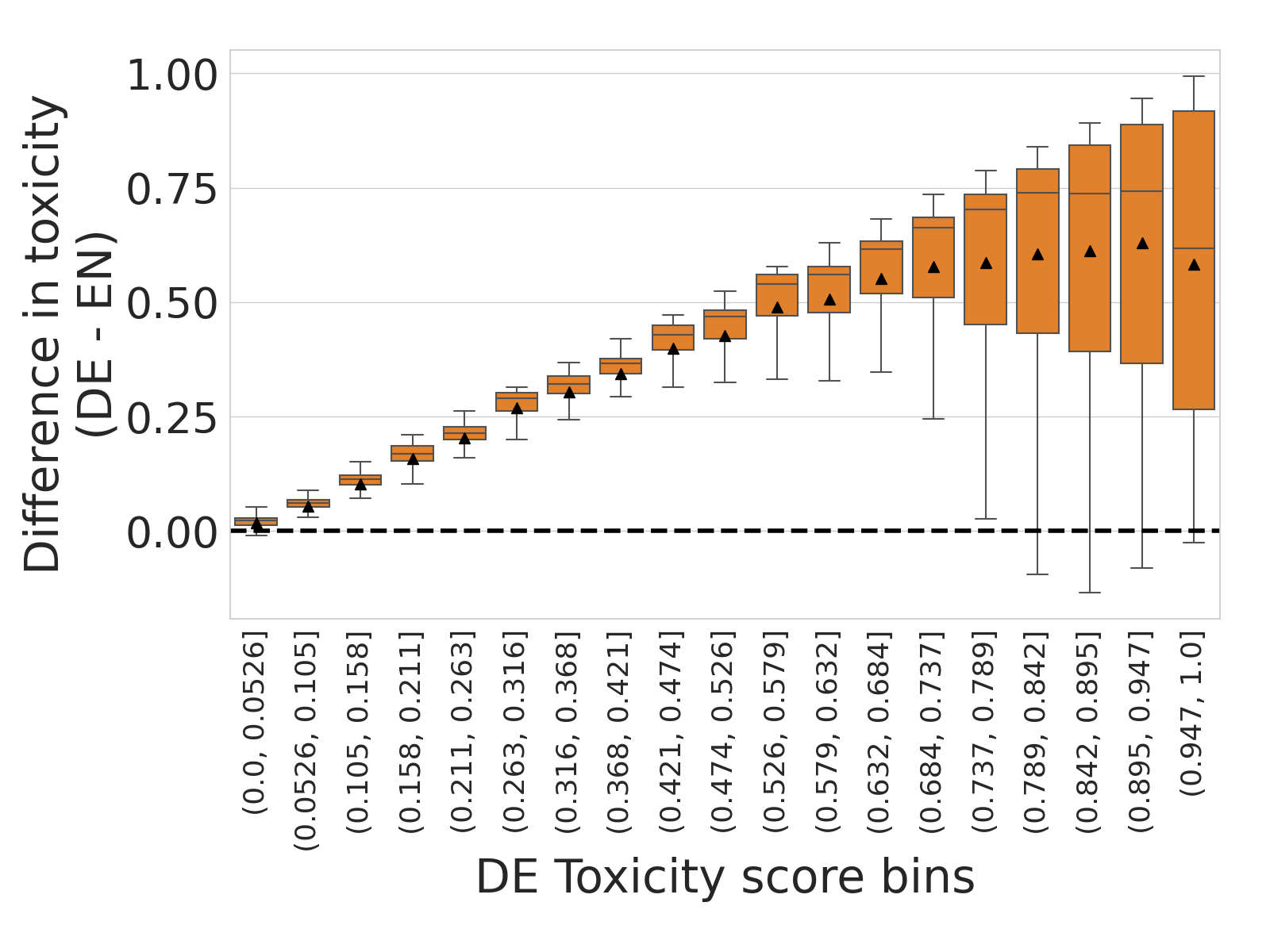}
\caption{Distributions of difference in toxicity between German tweets from Dataset 2 and their English translation, divided by 20 bins of toxicity computed on the German version of each tweet. Outliers are not shown and triangles indicate the mean of the distribution.}\label{fig:de_difference_versus_toxicity}
\end{figure}

We study how Perspective bias affects German tweets in the range of toxicity [0, 1] by dividing tweets of Dataset 2 in bins based on their German toxicity score, and computing the difference with the toxicity score assigned to their English translation, for each bin. Results are shown in Figure \ref{fig:de_difference_versus_toxicity} and highlight that large differences are found in tweets with high German toxicity scores. We also find a significant positive Pearson correlation (R=0.9, $P < .001$) between difference and toxicity (not binned). These results highlight that the bias of Perspective API is significantly higher for more toxic tweets. In other words, German tweets with a larger toxicity score are significantly over-estimated as toxic.

\subsubsection{Implications for content moderation}
To assess the implications of this bias, we analyze how moderation strategies aimed at minimizing online toxicity might affect decision-making processes, especially when a moderator or an automatic system relies on toxicity scores from the Perspective API to oversee and regulate discussions on a social media platform \cite{rieder2021fabrics}.

\begin{figure}[t!]
    \centering
\includegraphics[width=\linewidth]{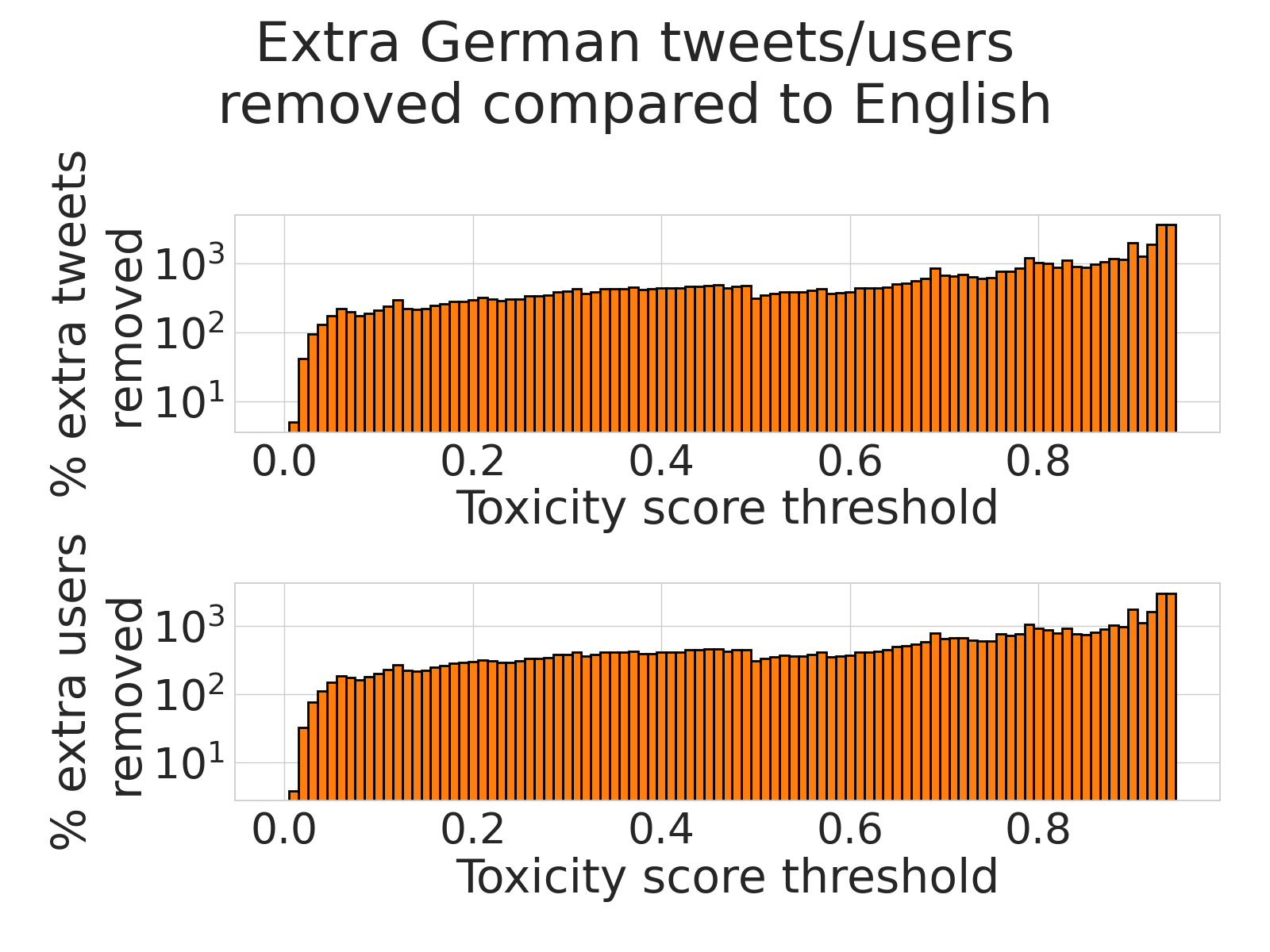}
\caption{Proportion of extra tweets (top) and users (bottom) removed in German compared to English for different choices of toxicity score as a threshold, i.e., tweets/users above a given score would be removed. The median increase in percentage points is +429.94 for tweets and +409.84 for users, meaning that on average more than 4 times German tweets and users would be removed compared to English.}\label{fig:de_it_removal}
\end{figure}

Specifically, we focus on how different toxicity thresholds used for removing tweets or users might affect COVID-19 vaccine conversations in German when compared to their translation in English, using the random sample of \num{10000} tweets employed in the previous analysis. 
In the top panel of Figure \ref{fig:de_it_removal}, we simulate removing all tweets whose toxicity score exceeds a given threshold, and we show the percentage of extra tweets removed when considering German scores compared to English scores. For instance, by adopting the threshold value of $0.7$ recommended by Perspective API\footnote{\url{https://developers.perspectiveapi.com/s/about-the-api-score?language=en_US}}, 10 times more German tweets would be removed, and similarly for the number of German users (as shown in the bottom panel of the figure). The median increase in percentage points is +429.94 for tweets and +409.84 for users, meaning that on average more than four times German tweets and users would be removed compared to English.

\subsection{Wikipedia texts in German are more toxic than in English}
\begin{figure}[t!]
    \centering
    \includegraphics[width=\linewidth]{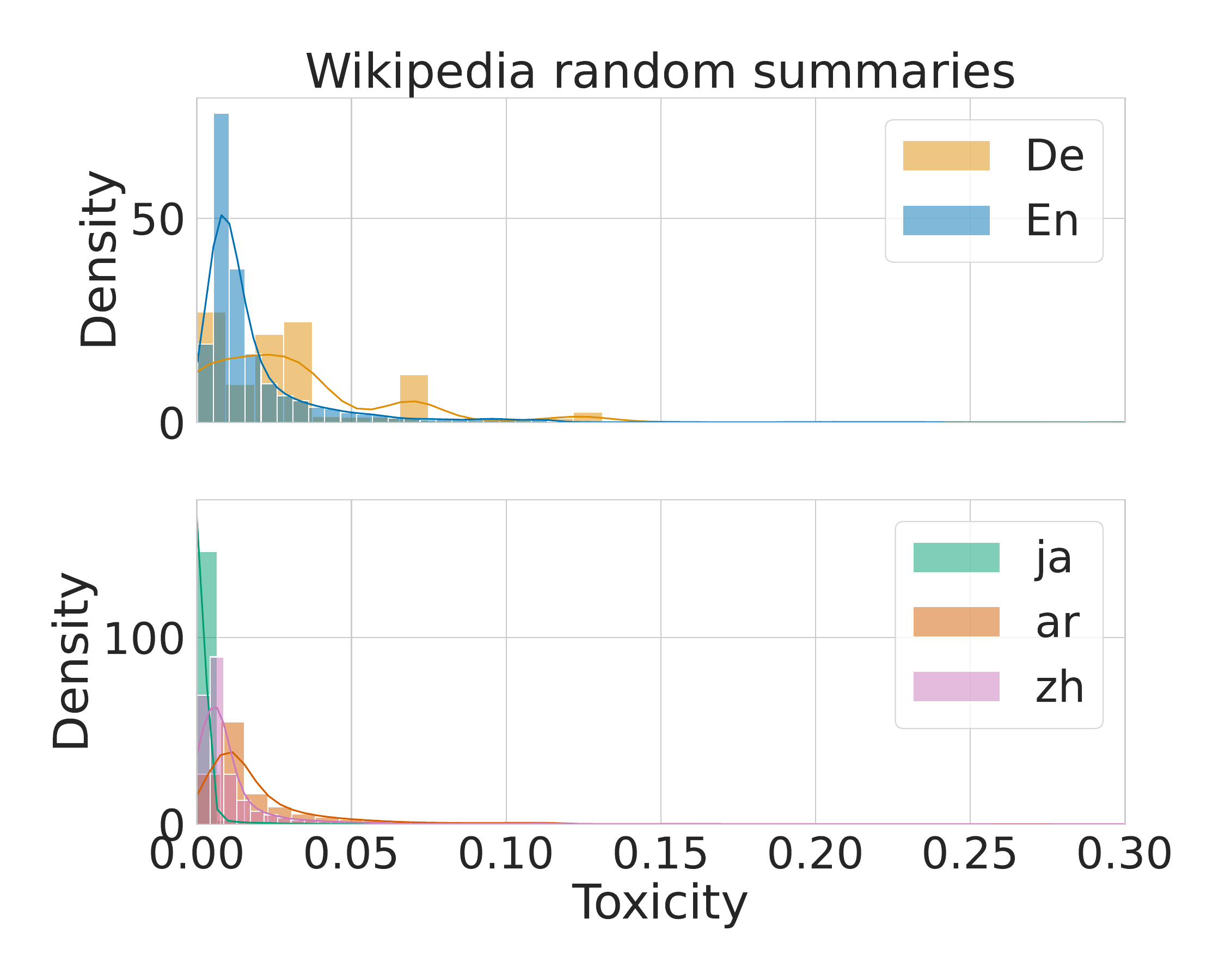}
    \caption{Distribution of toxicity for over \num{10000} random Wikipedia page summaries in Arabic (ar), Chinese (zh), English (en), German (de) and Japanese (ja). Histograms are built with 100 equal-width bins. We limit the x-axis to 0.3 for clarity reasons, as the distribution is highly skewed towards 0.}\label{fig:de_en_texts}
\end{figure}

To verify that the identified biases do not merely apply to Twitter messages, we also examine the distribution of toxicity scores when classifying a random sample of summaries from Wikipedia. As shown in the top panel of Figure \ref{fig:de_en_texts}, also for these texts, German exhibits significantly larger toxicity scores than those in English (Kruskal-Wallis $P < .001$). We also observe peaks in the distribution of German scores, which resemble those of previous distributions, such as Figure \ref{fig:de_eu_bias}.
We perform an additional robustness test by analyzing Wikipedia summaries in three non-European languages supported by Perspective API (Arabic, Chinese and Japanese) and computing their toxicity scores. As shown in the bottom panel of Figure \ref{fig:de_en_texts}, they do not exhibit abnormal peaks as for the German language. An analysis of the most frequent scores confirms the findings provided in Figure \ref{fig:de_eu_bias_digits}, with German texts exhibiting scores that are disproportionately more frequent than other languages.

\section{Discussion}
\subsection{Contributions}
Perspective API has emerged as the technological foundation for a broad spectrum of applications, stretching from academic research to practical implementations in content moderation across various online platforms. In the realm of research, scholars and data scientists utilize Perspective API to analyze online discourse, study patterns of hate speech and harassment, and explore the dynamics of digital communication. Beyond the academic sphere, Perspective API plays a pivotal role in content moderation for websites, forums, and large-scale social media platforms. By automatically flagging potentially harmful or abusive comments, it aids moderators in maintaining a constructive and respectful environment, enabling large-scale platforms to manage large volumes of user-generated content that would make manual moderation impractical. Perspective API is often integrated into comment systems and discussion forums, enabling preemptive identification and removal of messages that are classified as toxic. For these reasons, automated tools such as Perspective API apparently contribute to overcoming the challenges in content moderation and to respecting the regulatory obligations that large online platforms must abide by \cite{trujillo2023dsa}. However, while the widespread application of such tools means that these systems provide important support for researchers and platforms alike, it also highlights the potential negative impact of any biases or inaccuracies in the data and algorithms.

This paper identified a significant bias within the Perspective API model when it comes to processing German-language messages: our analysis reveals that the algorithm tends to assign disproportionately higher toxicity scores to texts written in German, compared to similar content in other languages. Specifically, our examination of Twitter posts revealed that the same content reported in German was more toxic than its English-translated version. This pattern was further validated by analyzing random texts from Wikipedia summaries in both languages. The distribution of toxicity returned by the API for German appears spiky, rather than following the smooth distribution of other languages -- suggesting some errors or artifacts stemming from the underlying machine learning model. Due to the black box and proprietary nature of the model, we unfortunately cannot examine the underlying reasons for these results, which calls for additional research efforts. However, an exploratory analysis showed that no specific characters and words are associated with such spiky behaviour, and that no abnormal patterns are observed for other non-European languages such as Arabic, Chinese and Japanese.

This bias raises profound concerns, as it not only skews research findings but also impacts the efficacy of content moderation on digital platforms. In the academic context, such a bias can lead to misleading conclusions about the nature and prevalence of toxicity within German-speaking online communities. For content moderators and platform administrators, this skewed assessment could result in unjust censorship or the inadvertent promotion of a hostile online environment for German-speaking users. For academic and journalistic research relying on the API, the implications are far-reaching, potentially affecting the perception and treatment of German digital discourse globally.

The problems identified in this paper highlight the inherent risks as researchers become increasingly dependent on proprietary models for data analysis and AI. The closed and proprietary APIs offered by major tech companies provide researchers with easy access to vast amounts of data and sophisticated analysis methods, which can otherwise be challenging to access. This ready availability is invaluable for studying complex social phenomena, trends, and behaviors in the digital age. However, the growing dependency on closed models also raises critical concerns. The opaque nature of these proprietary systems means that researchers are unable to scrutinize or understand the inner workings of the algorithms, and identify potential biases and limitations. The result is a  substantial risk of systemic biases being embedded and perpetuated through their employment. This lack of transparency can furthermore inadvertently influence the direction and conclusions of studies, as researchers must operate within the constraints and methodologies predefined by these tools. While proprietary APIs have opened new avenues for exploring social sciences, they also necessitate a cautious and critical approach to ensure the integrity of research and its reproducibility \cite{pozzobon2023challenges} -- rather than taking the technology firms by their word.

\subsection{Limitations}
There are some limitations in our work. We based our investigation on data from Twitter, which is not the most widely adopted social platform in many countries, and therefore not fully representative of the population. However, our robustness analyses with Wikipedia data suggest that the highlighted issues do not depend on user demographics  or the specific context/topic of conversation. Besides, we only analyzed the Toxicity attribute of the API, and leave an investigation of other attributes (such as Insult or Profanity) for future research. Lastly, We did not evaluate whether this bias manifests in other LLM-based models for toxicity detection.

\subsection{Ethical considerations}
We envision positive ethical implications for this work, particularly concerning future research on online toxicity as well as on the policy design of online platforms. Our findings underscore the potential consequences of relying on black-box models and APIs, without a thorough understanding of their language biases. The identification of intrinsic biases within the multilingual Perspective API highlights the risk of inadvertently amplifying or suppressing certain voices in online spaces. The observed higher levels of toxicity assigned to German content compared to other languages raise concerns about the fairness and equity of ongoing content moderation processes, which could potentially lead to disproportionate censorship or scrutiny of specific linguistic communities. This study urges a reevaluation of the ethical dimensions surrounding the deployment of such tools, emphasizing the need for transparency, accountability, and ongoing refinement to ensure that AI-driven content moderation aligns with principles of impartiality and cultural sensitivity, thereby mitigating the inadvertent amplification of biases in online discourse.

\subsection{Data availability}
To ensure reproducibility, we will provide access to the code and datasets employed in our analysis in a public repository after the revision process. For what concerns Dataset 1 and Dataset 2, to comply with Twitter's Terms of Service, we can only provide tweet IDs, which can be employed (\textit{hydrated}) to retrieve the data by means of Twitter API. We acknowledge that access to the \texttt{GET /2/tweets} endpoint for researchers is not as easy as in the past, and we encourage interested researchers to reach out to the authors in case they encounter difficulties in replicating the analyses. For what concerns Dataset 3, we will provide the full data, as it is already publicly accessible. We notice that we do not attempt to de-identify or disclose any personal or sensitive data.

\section{Appendix}
Here we provide supplementary figures and results referenced and commented in the main text.

\begin{figure}[!t]
    \centering
    \includegraphics[width=\linewidth]{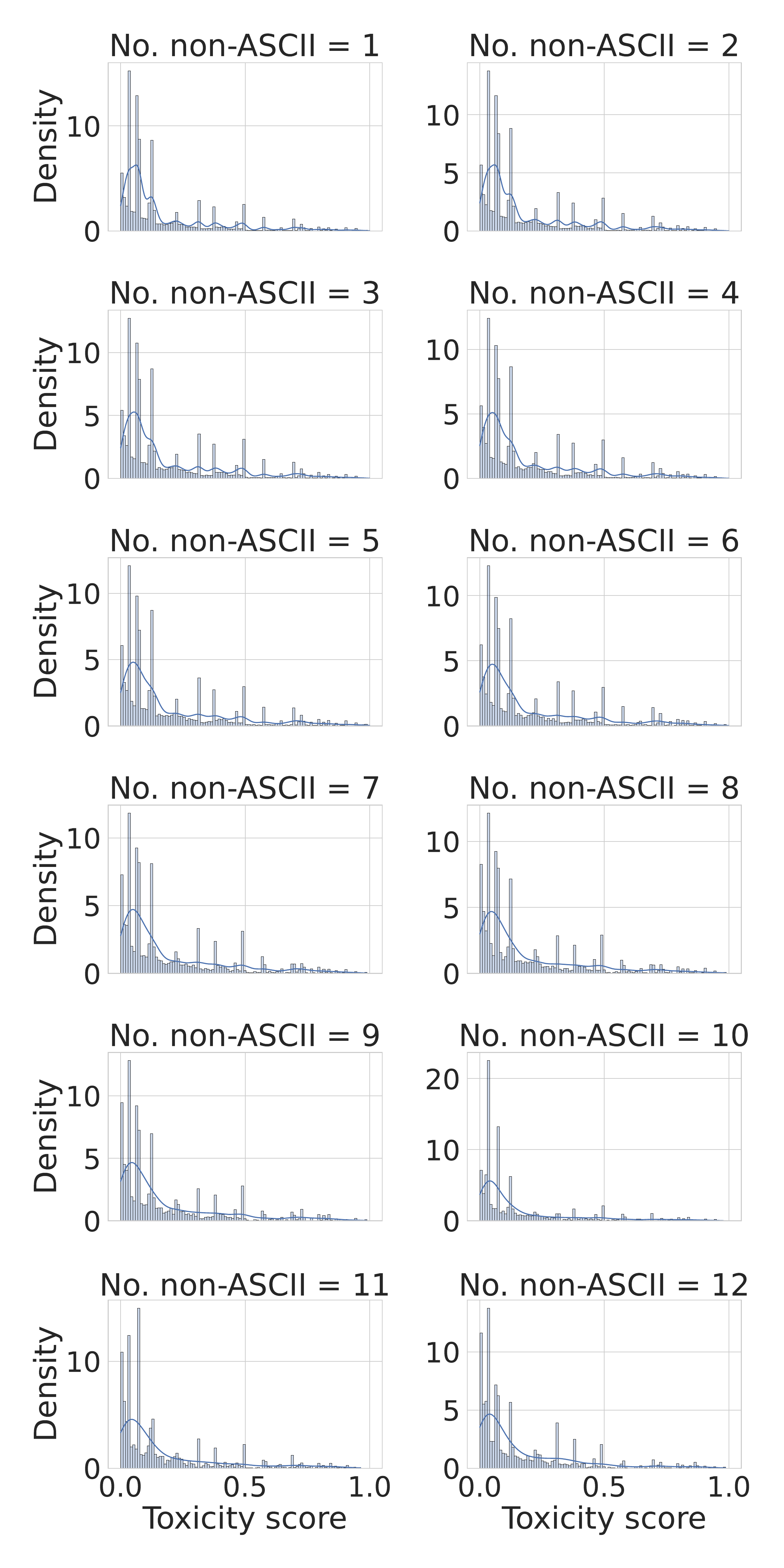}
    \caption{Distributions of toxicity scores for tweets from German-speaking countries in Dataset 1 based on the number of non-ASCII characters they contain.}
\label{fig:de_ascii_appendix}
\end{figure}

Figure \ref{fig:de_ascii_appendix} provides the distribution of toxicity scores for tweets shared in German-speaking countries from Dataset 1 separating them by the number of non-ASCII characters they contain.

\begin{figure}[!t]
    \centering
\includegraphics[width=\linewidth]{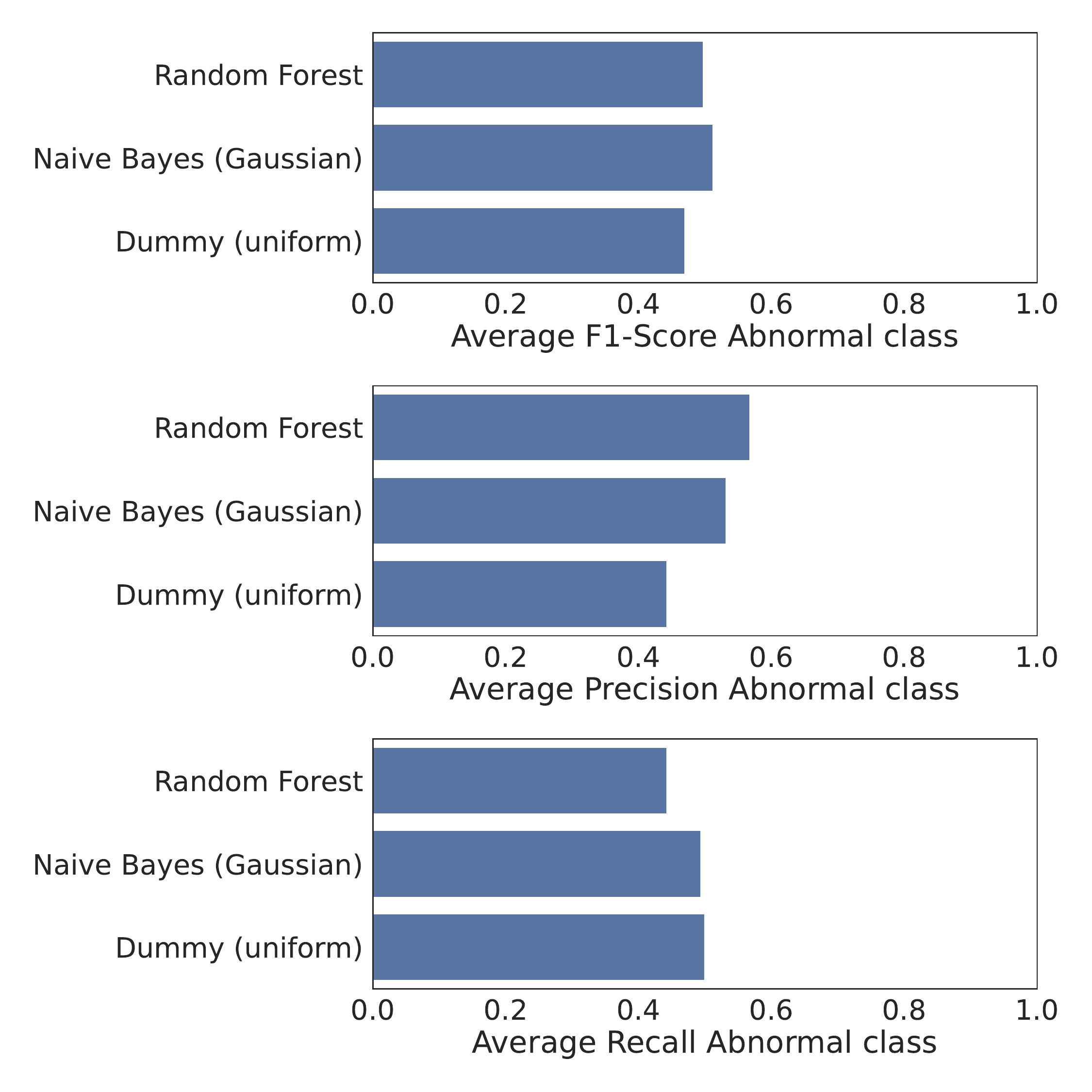}
    \caption{Performance evaluation of three classifiers in the binary task of predicting whether a tweet belongs to the class of Abnormal toxicity scores described in the main text. We show the mean value of each metric in a 5-fold cross-validation.}
\label{fig:classification}
\end{figure}

Figure \ref{fig:classification} provides the classification performance, computed with a 5-fold cross-validation, of three off-the-shelf classifiers in the binary task of predicting whether a tweet belongs to the Abnormal class described in the main text.

\begin{figure}[!t]
    \centering
\includegraphics[width=\linewidth]{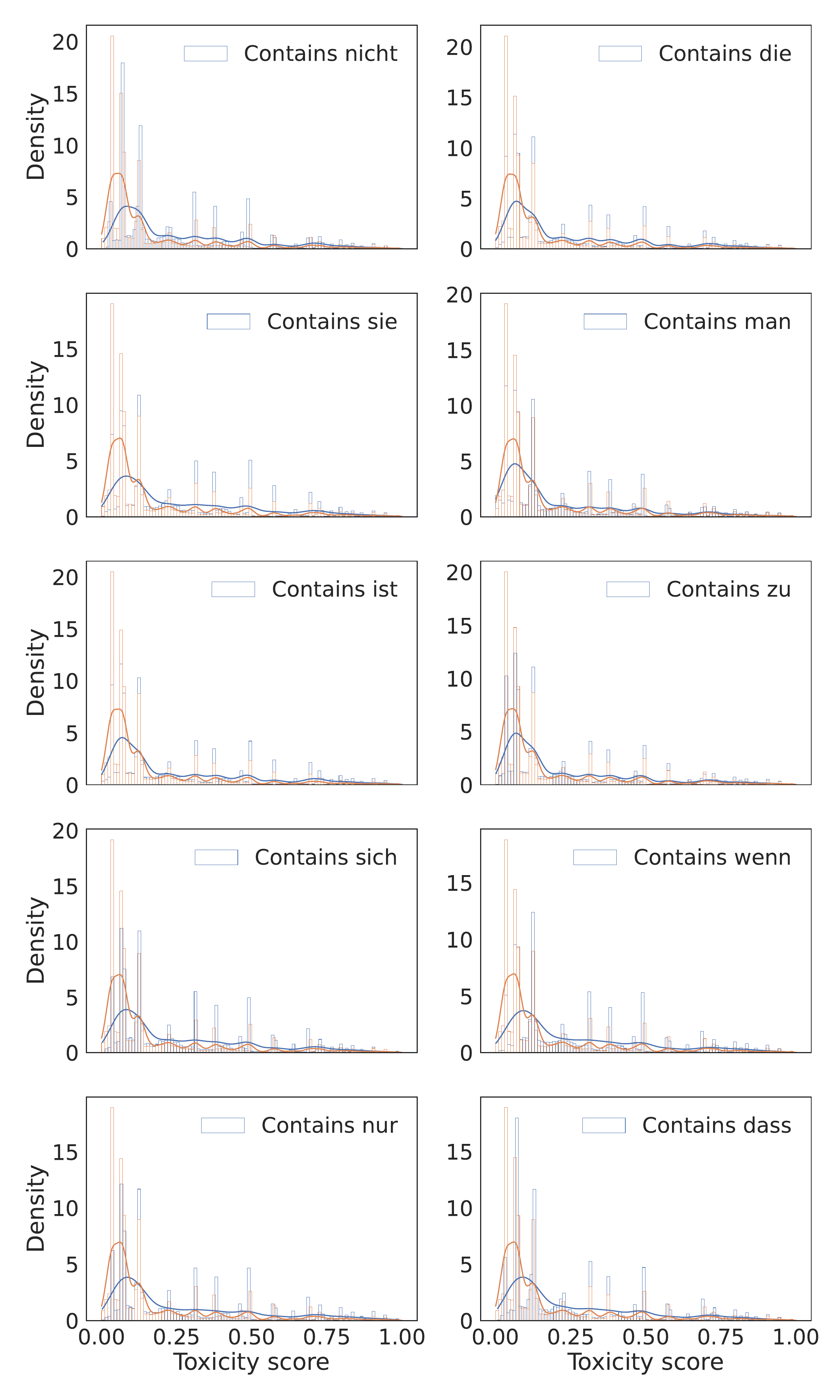}
    \caption{Distributions of toxicity scores for tweets from German-speaking countries in Dataset 1 that contain (or not) one of the top-10 words differentiating the Abnormal class.}
\label{fig:fighting_words}
\end{figure}

Figure \ref{fig:fighting_words} provides the distributions of toxicity scores for tweets containing (or not) one of the top 10 words differentiating the Abnormal class (in terms of Z-score). They are translated into English as follows:
\begin{itemize}
    \item nicht: not
    \item die: the
    \item sie: they/she
    \item man: one
    \item ist: is
    \item zu: to
    \item sich: oneself
    \item wenn: when
    \item nur: only
    \item dass: that
\end{itemize}

\bibliographystyle{abbrv}
\bibliography{references}

\end{document}